# The second law of blackhole dynamics

**Koustubh Ajit Kabe**
Department of Physics, S.S. & L.S. Patkar College of Arts & Science, S.V. Road
Goregaon (West) Mumbai - 400 062 INDIA.

**ABSTRACT**

In this paper, the second law blackhole dynamics is restated, with a rigid proof, in a different form akin to the statement of the second law of thermodynamics given by Clausius. The various physical possibilities and implications of this statement are discussed therein.

It is observed that dynamics of blackholes runs competitively on the same lines as that of classical thermodynamics, so that the former becomes a subject in itself. This subject is now referred to as *blackhole dynamics*. The theory of blackhole dynamics was arrived at due to the pioneering works of Bekenstein [1, 2], Hawking [3], Carter, Bardeen & others [4]. The so-called *No-Hair Theorem* (NHT) was enunciated by J. A. Wheeler and workable proofs, part wise were provided by Hawking, Carter, Bardeen & Israel [5] which only when integrated would account for the NHT. Just as the four laws of thermodynamics, there are the four laws of blackhole dynamics (for a proper well integrated, coherent & comprehensive reference to these laws see [6] & related references cited therein).

The following paper looks up at an alternate statement of the second law of blackhole dynamics and its effect on the Hawking radiation. The most general form of the second law is due to Hawking, and explicitly states that in any classical interaction of matter and radiation with blackholes, the total surface area of the boundaries of these holes (as formed by their horizons) can never decrease. Thus, for coalescing blackholes with $\partial B_i$ as their boundaries $(i = 1,2,3,...,n)$ and $A(\partial B_i)$ as their corresponding areas,

$$\sum_{i=1}^{n} A(\partial B_i) \leq A(\partial B), \qquad (1)$$

where $\partial B_i$ is the boundary of the blackhole formed from the coalescence of the blackholes.

The energy is conserved for the blackholes as well as it is for the other phenomena that occur in the universe. This is embodied in the first law of blackhole dynamics;

$$d(Mc^2) = \frac{\kappa c^2}{8\pi G} dA + \Omega dJ + \varphi dQ \qquad (2)$$

where the symbols carry their usual meaning in accord with the conventions of blackhole dynamics. Here, we see that surface gravity corresponds to the temperature and area to the entropy. Within the classical framework, proportionality factors are undetermined. The Hawking radiation suggests that one should associate a temperature and entropy to a blackhole given by

$$T_H = \frac{\kappa \hbar}{2\pi k_B c} \qquad (3)$$

and

$$S_{BH} = \frac{k_B}{4 l_{Pl}^2} A \qquad (4)$$

where $k_B$ is the Boltzmann, constant. $S_{BH}$ is the so-called Bekenstein-Hawking entropy and $T_H$ is the Hawking temperature.

We base this paper on the statement of the second law of thermodynamics as given by Clausius,

stated explicitly as: *it is impossible for a self acting machine, unaided by any external agency to convey heat from a body at a lower temperature to a body at a higher temperature.*
The self acting machine as well as the unaided part therein can be omitted (the blackhole in itself is self acting and unaided) since the most energetic phenomena in the universe are involved in blackhole physics and phenomenology, and hence no aid can be conceived and/ or be provided to any concerned blackhole: thus the aid is from the blackhole itself. We state a similar form of the second law for blackhole dynamics and then give the proof of it and then restate the law in various different forms.

*Statement of the second law:*

No information equivalent to radiation (in particular, Hawking radiation) can be communicated by a blackhole to a blackhole with a relatively higher value of surface gravity.

*Proof of the second law:*

Let $A_1 \equiv A(\partial B_1)$ be the area of the blackhole whose boundary is $\partial B_1$ and to this we assign the surface gravity $\kappa_1 \equiv \kappa(\partial B_1)$, similarly $A_2 \equiv A(\partial B_2)$ is the area of the blackhole $\partial B_2$ with surface gravity $\kappa_2 \equiv \kappa(\partial B_2)$. Let $\partial B_1$ communicate a radiation by the Hawking mechanism to $\partial B_2$ subject to the assumed subsidiary condition: $\kappa_1 < \kappa_2$. The energy equivalent of the radiation is $E = mc^2 = \hbar\omega$.
The blackhole $\partial B_1$ will then experience a decrease in its area $A_1$, as

$$A_1' = A_1 - \delta A \tag{5}$$

and correspondingly, $\partial B_2$ will experience an areal increment, as

$$A_2' = A_2 + \delta A. \tag{6}$$

now, the temperature of a blackhole called its Hawking temperature is given by eq (3) and to a first approximation in eq (2), take $\Omega = \varphi = 0$ and/ or alternately take $J = Q = const.$ and we now have an inverse dependence of horizon area on the surface gravity which is true in general also i.e. without taking $\Omega = \varphi = 0$ or $J = Q = const.$ (the null and constancy assumption have been made to get a simpler mathematical picture and to avoid large expressions involving the *J* and *Q* terms in the proof given below), as

$$dA = \frac{8\pi \, dM}{\kappa}. \tag{7}$$

In the case of the blackholes $\partial B_1$ and $\partial B_2$,

$$S_2' > S_1' \Rightarrow A_2' > A_1', \tag{8}$$

that is

$$\frac{dM_2}{\kappa_2} > \frac{dM_1}{\kappa_1} \tag{9}$$

or

$$\kappa_2 < \kappa_1 \tag{10}$$

which is clearly a contradiction to our earlier assumed subsidiary condition: $\kappa_1 < \kappa_2$.
Our newly enacted blackhole scenario is now theoretically true. Thus, if a blackhole emits Hawking radiation, the emitted particle cannot be trapped by the gravity of any other blackhole with a higher value of surface gravity. Else the particle would by some mechanism, without violating the law of conservation of energy, tunnel through the gravitational potential barrier and escape and simultaneously get red-shifted. If the blackhole with the higher value of surface gravity is rotating Kerr or Kerr-Newman blackhole then the particle can take a partner with it and escape by the *Penrose Mechanism* or by the *Blandford-Znacjek*. This could provide a better solution to the problem of the *information paradox* of Hawking: details of this will be found elsewhere. If the particle escapes from the blackhole with a higher value of surface gravity, say $\kappa$, then, the new surface gravity would be $\kappa_0$, and

$$\frac{(\kappa - \kappa_0)A}{c^2} = const.$$

where $A \equiv A(\partial B)$ is the area of the event horizon of the blackhole that we have considered here. The constant corresponds to the new characteristic wavelength of the red-shifted escaping particle. Thus, every particle that had event horizon as its history is always connected with blackholes of higher value of surface gravity than that of its horizon and it is possible that all the blackholes are connected in a higher dimensional physics (See for example [7] and references therein). The particle will be accelerated or decelerated by a factor of $g = \kappa \sim \kappa_0$ according as the factor is negative or positive.

Another way of stating our law would then be a combination of the original areal law and the surface gravity law as:

1.  *In any interaction involving blackholes, the area of the boundary of a blackhole with a higher value of surface gravity cannot increase at the cost of decrease in the corresponding area of any other blackhole in the universe with a lower value of surface gravity.*

Or

2.  *A blackhole cannot trap a particle emitted by another blackhole having a relatively lower value of surface gravity, by the Hawking mechanism or any other method.*

Or

3.  *A blackhole with a higher value of surface gravity has a zero capture cross-section with respect to the radiation emitted by a blackhole with a relatively lower value of surface gravity.*

The blackholes could thus possibly have a non-trivial topology of a multiply connected domain in higher dimension with some associated constant of action, a quantum rule and an associated conservation law and the corresponding conserved quantity (in higher dimension). Only then is it possible for a blackhole with a higher value of surface gravity to reject information/ energy communicated to it by a blackhole with a lower value of surface gravity. The work of Horowitz and Polchinsky [7] is a notable feature of the higher dimensional connection that has here been extended to multiple connections without any reference to strings. Conventionally, one may also interpret this as a negative energy communicated in positive direction in spacetime reinterpreted as positive energy communicated backwards in time. One of the various tunneling mechanisms have been considered by the author [8] and is history dependent.

__________________________________________